\documentclass[reqno]{amsart}

\usepackage[hyphens]{url}
\usepackage[style=numeric,sorting=none,backend=biber,maxbibnames=99]{biblatex}
\usepackage{tiz-note}
\bibliography{main}
\usepackage{tiz-paper}

\usepackage{newunicodechar}
\newunicodechar{≥}{\ensuremath{\mathnormal\geq}}
\newunicodechar{Π}{\ensuremath{\mathrm\Pi}}
\newunicodechar{Δ}{\ensuremath{\mathnormal\Delta}}
\newunicodechar{δ}{\ensuremath{\mathnormal\delta}}
\newunicodechar{Θ}{\ensuremath{\mathnormal\Theta}}
\newunicodechar{θ}{\ensuremath{\mathnormal\theta}}
\newunicodechar{σ}{\ensuremath{\mathnormal\sigma}}
\newunicodechar{ν}{\ensuremath{\mathnormal\nu}}
\newunicodechar{Ξ}{\ensuremath{\mathnormal\Xi}}
\newunicodechar{η}{\ensuremath{\mathnormal\eta}}
\newunicodechar{β}{\ensuremath{\mathnormal\beta}}
\newunicodechar{ε}{\ensuremath{\mathnormal\epsilon}}
\newunicodechar{▷}{\ensuremath{\mathnormal\vartriangleright}}
\newunicodechar{π}{\ensuremath{\mathnormal\pi}}
\newunicodechar{Γ}{\ensuremath{\mathnormal\Gamma}}
\newunicodechar{•}{\ensuremath{\mathnormal\bullet}}
\newunicodechar{∘}{\ensuremath{\mathnormal\circ}}
\newunicodechar{₁}{\ensuremath{\mathnormal{_1}}}
\newunicodechar{₂}{\ensuremath{\mathnormal{_2}}}
\newunicodechar{₃}{\ensuremath{\mathnormal{_3}}}
\newunicodechar{₄}{\ensuremath{\mathnormal{_4}}}

\DeclareUnicodeCharacter{0308}{ErrorErrorErrorErrorErrorErrorErrorErrorErrorErrorErrorError}

\newunicodechar{∀}{\ensuremath{\mathnormal\forall}}
\newunicodechar{λ}{\ensuremath{\mathnormal\lambda}}
\newunicodechar{∷}{\ensuremath{\mathnormal{::}}}
\newunicodechar{×}{\ensuremath{\mathnormal\times}}
\newunicodechar{Σ}{\ensuremath{\mathnormal\Sigma}}
\newunicodechar{←}{\ensuremath{\mathnormal\from}}
\newunicodechar{→}{\ensuremath{\mathnormal\to}}
\newunicodechar{≡}{\ensuremath{\mathrel{\equiv}}}
\newunicodechar{≃}{\ensuremath{\mathrel{\simeq}}}
\newunicodechar{ℓ}{\ensuremath{\mathrel{\ell}}}
\newunicodechar{⊎}{\ensuremath{\mathrel{\uplus}}}
\newunicodechar{ℕ}{\ensuremath{\mathnormal{\mathbb N}}}
\newunicodechar{ℤ}{\ensuremath{\mathnormal{\mathbb Z}}}
\newunicodechar{¹}{\ensuremath{\mathnormal{^1}}}
\newunicodechar{₀}{\ensuremath{\mathnormal{_0}}}
\newunicodechar{⊤}{\ensuremath{\mathnormal{\top}}}
\newunicodechar{∞}{\ensuremath{\mathnormal{\infty}}}

\definecolor{AyaFn}{HTML}{00627a}
\definecolor{AyaConstructor}{HTML}{067d17}
\definecolor{AyaStruct}{HTML}{00627a}
\definecolor{AyaGeneralized}{HTML}{00627a}
\definecolor{AyaData}{HTML}{00627a}
\definecolor{AyaPrimitive}{HTML}{00627a}
\definecolor{AyaKeyword}{HTML}{0033b3}
\definecolor{AyaComment}{HTML}{8c8c8c}
\definecolor{AyaField}{HTML}{871094}
\newcommand\AyaFn[1]{\textcolor{AyaFn}{#1}}
\newcommand\AyaConstructor[1]{\textcolor{AyaConstructor}{#1}}

\newcommand\AyaGeneralized[1]{\textcolor{AyaGeneralized}{#1}}
\newcommand\AyaData[1]{\textcolor{AyaData}{#1}}
\newcommand\AyaPrimitive[1]{\textcolor{AyaPrimitive}{#1}}
\newcommand\AyaKeyword[1]{\textcolor{AyaKeyword}{#1}}
\newcommand\AyaLocalVar[1]{\textit{#1}}

\pgfdeclarelayer{frontmost}
\pgfsetlayers{main,frontmost}
\usetikzlibrary{patterns}

\tikzset{
  carlo-axes/.style =
  {
    y = {(0,-1)},
    z = {(-0.6,0.6)}
  } ,
  shorten <>/.style =
  {
    shorten >=#1 , shorten <=#1
  } ,
  equals arrow/.style =
  {
    arrows = - ,
    double equal sign distance ,
  } ,
}
\newcommand{\carloTikZ}[1]{
  \begin{tikzpicture}[carlo-axes, scale = 1.6]
		#1
	\end{tikzpicture}}
\newcommand{\carloCoordSq}[2]{\begin{scope}[shift={(-1.5,0)},scale=0.4]
    \draw [->] (0,0) to node [above,pos=1] {#1} (1,0) ;
    \draw [->] (0,0) to node [right,pos=1] {#2} (0,1) ;
  \end{scope}}
\newcommand{\carloCTikZ}[1]{\begin{center}
\carloTikZ{#1}
\end{center}}

\usepackage[hyphenbreaks]{breakurl}

\begin{document}

\title{Two tricks to trivialize higher-indexed families}

\author{Tesla Zhang}
\address{Carnegie Mellon University}
\email{teslaz@cmu.edu}
\date\today

\begin{abstract}
The conventional general syntax of indexed families in dependent type theories follow the style of
``constructors returning a special case'', as in Agda, Lean, Idris, Coq, and probably many other systems.
\textit{Fording} is a method to encode indexed families of this style
with index-free inductive types and an identity type.

There is another trick that merges interleaved higher inductive-inductive types into a single
big family of types. It makes use of a small universe as the index to distinguish the original types.

In this paper, we show that these two methods can trivialize some very fancy-looking indexed families
with higher inductive indices (which we refer to as \textit{higher indexed families}).
\end{abstract}
\maketitle
\tableofcontents
In this article, we show that the idea of \textit{Fording} proposed by Conor McBride in~\cite{GAL}
makes the ``mini-universe'' definition of integers $\mathbb{Z}_U$ in~\cite{Integers} trivial.

\begin{warning}
This is a very syntax-minded article, so there will be pseudocode snippets
to provide intuition based on imaginary syntax, but those will be mostly
analogous to the style in the HoTT book~\cite{HoTTBook} and Cubical Agda~\cite{CubicalAgda}.
\end{warning}

We will also use the Aya language,
which adapts a ``pattern matching'' syntax~\cite{SIT} for indexed families.
For instance, the traditional definition of sized vectors looks like the following:

{
\vspace{0.15cm}
\RaggedRight
\setlength\parindent{0pt}
\setlength{\leftskip}{1cm}
\AyaKeyword{open}\hspace{0.5em}\AyaKeyword{data}\hspace{0.5em}\AyaData{Vec}\hspace{0.5em}\((\)\AyaLocalVar{A}\hspace{0.5em}\(:\)\hspace{0.5em}\AyaKeyword{Type}\()\)\hspace{0.5em}\((\)\AyaLocalVar{n}\hspace{0.5em}\(:\)\hspace{0.5em}\AyaData{Nat}\()\)~\\
\(\mid\)\hspace{0.5em}\AyaLocalVar{A},\hspace{0.5em}\AyaConstructor{zero}\hspace{0.5em}\(\Rightarrow\)\hspace{0.5em}\AyaConstructor{nil}~\\
\(\mid\)\hspace{0.5em}\AyaLocalVar{A},\hspace{0.5em}\AyaConstructor{suc}\hspace{0.5em}\AyaLocalVar{n}\hspace{0.5em}\(\Rightarrow\)\hspace{0.5em}\AyaConstructor{cons}\hspace{0.5em}\AyaLocalVar{A}\hspace{0.5em}\((\)\AyaData{Vec}\hspace{0.5em}\AyaLocalVar{A}\hspace{0.5em}\AyaLocalVar{n}\()\)

\setlength{\leftskip}{0cm}
\vspace{0.15cm}
}

It says that ``the type \AyaData{Vec} has two constructors, \AyaConstructor{nil}
is only available when the second argument of \AyaData{Vec} is \AyaConstructor{zero},
and similarly for \AyaConstructor{cons} but only when it's $\AyaConstructor{\mathrm{suc}}~n$''.
The arguments of the constructors are listed like in Haskell,
and unlike the GADT syntax adapted in Haskell GADT~\cite{GADTHS},
Agda~\cite{Agda}, and Idris, we do not write out the return type
of constructors.

This \textit{pattern matching} does not need to be exhaustive and can have \textit{redundant} cases,
because they are actually not pattern matching, but instead describes the \textit{availability}
of constructors.
This syntax enjoys many benefits compared to the traditional one, discussed in~\cite{SIT}.

\section{Motivation}

The definition of $\mathbb{Z}_U$ is said to be inspired from a trick due to Szumi Xie\footnote{
Used in~\url{https://bitbucket.org/szumixie/tt-in-tt/src/master/Cubical/Syntax.agda},
a formalization of type theory syntax in Cubical Agda~\cite{CubicalAgda}},
which is also seen in the Haskell parser generator BNFC\footnote{
Website: \url{https://bnfc.digitalgrammars.com}, GitHub: \url{https://github.com/BNFC/bnfc}},
where the \texttt{-{}-haskell-gadt} option generates a big GADT that combines several data types
in the following way:
\begin{enumerate}
  \item Define an ``enumeration'' $U$, say, a data type with all constructors having no argument,
  \item For every data type $D$, add a constructor $D_U$ to $U$,
  \item Define a GADT $T$ indexed by $U$, say, $T:U\to\mathcal U$,
  \item For every constructor $c:X \to D$, add the following constructor to $T$:
  \[
    c_T : X \to T~D_U
  \]
\end{enumerate}
\begin{example}
Suppose we have the following data types,
written in Haskell:

{
\vspace{0.15cm}
\RaggedRight
\setlength\parindent{0pt}
\setlength{\leftskip}{1cm}
\AyaKeyword{data}\hspace{0.5em}\AyaData{D1}\hspace{0.5em}=\hspace{0.5em}\AyaConstructor{con₁}\hspace{0.5em}\AyaData{Int}\hspace{0.5em}\(\mid\)\hspace{0.5em}\AyaConstructor{con₂}\hspace{0.5em}\AyaData{Bool}~\\
\AyaKeyword{data}\hspace{0.5em}\AyaData{D2}\hspace{0.5em}=\hspace{0.5em}\AyaConstructor{con₃}\hspace{0.5em}\AyaData{Bool}\hspace{0.5em}\(\mid\)\hspace{0.5em}\AyaConstructor{con₄}\hspace{0.5em}\AyaData{Int}

\setlength{\leftskip}{0cm}
\vspace{0.15cm}
}
By applying the trick, we get the following GADT:

{
\vspace{0.15cm}
\RaggedRight
\setlength\parindent{0pt}
\setlength{\leftskip}{1cm}
\AyaKeyword{data}\hspace{0.5em}\AyaData{U}\hspace{0.5em}=\hspace{0.5em}\AyaConstructor{D1\textunderscore{}U}\hspace{0.5em}\(\mid\)\hspace{0.5em}\AyaConstructor{D2\textunderscore{}U}~\\
\AyaKeyword{data}\hspace{0.5em}\AyaData{T}\hspace{0.5em}:\hspace{0.5em}\((\)\AyaLocalVar{u}\hspace{0.5em}:\hspace{0.5em}\AyaData{U}\()\)\hspace{0.5em}\(\to\)\hspace{0.5em}\AyaKeyword{Type}\hspace{0.5em}\AyaKeyword{where}~\\
\hspace{1.0em}\AyaConstructor{con₁\textunderscore{}T}\hspace{0.5em}:\hspace{0.5em}\AyaData{Int}\hspace{0.5em}\(\to\)\hspace{0.5em}\AyaData{T}\hspace{0.5em}\AyaConstructor{D1\textunderscore{}U}~\\
\hspace{1.0em}\AyaConstructor{con₂\textunderscore{}T}\hspace{0.5em}:\hspace{0.5em}\AyaData{Bool}\hspace{0.5em}\(\to\)\hspace{0.5em}\AyaData{T}\hspace{0.5em}\AyaConstructor{D1\textunderscore{}U}~\\
\hspace{1.0em}\AyaConstructor{con₃\textunderscore{}T}\hspace{0.5em}:\hspace{0.5em}\AyaData{Bool}\hspace{0.5em}\(\to\)\hspace{0.5em}\AyaData{T}\hspace{0.5em}\AyaConstructor{D2\textunderscore{}U}~\\
\hspace{1.0em}\AyaConstructor{con₄\textunderscore{}T}\hspace{0.5em}:\hspace{0.5em}\AyaData{Int}\hspace{0.5em}\(\to\)\hspace{0.5em}\AyaData{T}\hspace{0.5em}\AyaConstructor{D2\textunderscore{}U}

\setlength{\leftskip}{0cm}
\vspace{0.15cm}
}
\end{example}
Szumi Xie's trick is essentially the same thing but applied to higher inductive-inductive types,
so we can encode interleaved recursion among constructors with an indexed higher inductive type.
This pattern is very common in the formalization of dependent type theory itself in dependent type theory.

So, relating this inspired definition with something else could
potentially open up a broader view of the trick, especially in homotopy type theory
or dependent type theories with quotient inductive types~\cite{QIIT}.

We start from an explanation of $\mathbb{Z}_U$ assuming familiarity with basic ideas
in homotopy type theory~\cite{HoTTBook}, including univalence,
the notion of equivalence (a function with the property of being an equivalence),
and higher inductive types (chapter 6 of~\cite{HoTTBook}).
These will be used as the context of discussion, but we will not go into details,
so even a vague impression should suffice.
Then we explain the Fording technique, and show how to Ford $\mathbb{Z}_U$.

\section{The peculiar definition of integers}

In~\cite{Integers}, Thorsten Altenkirch and Luis Scoccola
proposed the following definition of integers using a ``mini-universe'':

{
\vspace{0.15cm}
\RaggedRight
\setlength\parindent{0pt}
\setlength{\leftskip}{1cm}
\AyaKeyword{open}\hspace{0.5em}\AyaKeyword{data}\hspace{0.5em}\AyaData{U}~\\
\(\mid\)\hspace{0.5em}\AyaConstructor{T}~\\
\(\mid\)\hspace{0.5em}\AyaConstructor{path}\hspace{0.5em}\(:\)\hspace{0.5em}\AyaConstructor{T}\hspace{0.5em}\AyaFn{=}\hspace{0.5em}\AyaConstructor{T}~\\
~\\
\AyaKeyword{open}\hspace{0.5em}\AyaKeyword{data}\hspace{0.5em}\AyaData{PreInt}\hspace{0.5em}\AyaData{U}~\\
\(\mid\)\hspace{0.5em}\AyaConstructor{T}\hspace{0.5em}\(\Rightarrow\)\hspace{0.5em}\AyaConstructor{zero}~\\
~\\
\AyaKeyword{def}\hspace{0.5em}\AyaFn{Int}\hspace{0.5em}\(\Rightarrow\)\hspace{0.5em}\AyaData{PreInt}\hspace{0.5em}\AyaConstructor{T}

\setlength{\leftskip}{0cm}
\vspace{0.15cm}
}

They call it $\mathbb{Z}_U$, but we stick to an ASCII name for a controlled
use of Unicode characters in program source code.
This definition is not expected to the understood just by looking at it --
the rest of this section is devoted to provide the relevant background.

\begin{remark}
In the above snippet, \AyaData{PreInt} is a \textit{simpler indexed type}~\cite{SIT}.
It is equivalent to the following Agda-style constructor:
\[
  \AyaConstructor{\mathrm{zero}} : \AyaData{\mathrm{PreInt}}~\AyaConstructor{\mathrm{T}}
\]
\end{remark}

To decipher this definition, we need to understand what integers are.
Instead of just defining the sets (i.e. list its elements)
or the \textit{normal form} of integers, we look at the algebraic structure.

We think of them as a modified version of natural numbers,
where the successor function is not just a constructor (which is an injective map),
but is also equipped with a ``canonical inverse'' (so it's now bijective),
so it becomes an equivalence rather than a function.

So, it is tempting to write down the following definition:

{
\vspace{0.15cm}
\RaggedRight
\setlength\parindent{0pt}
\setlength{\leftskip}{1cm}
\AyaKeyword{data}\hspace{0.5em}\AyaData{Int}\hspace{0.5em}\AyaKeyword{where}~\\
\hspace{1.0em}\AyaConstructor{zero}\hspace{0.5em}:\hspace{0.5em}\AyaData{Int}~\\
\hspace{1.0em}\AyaConstructor{succ}\hspace{0.5em}:\hspace{0.5em}\AyaData{Int}\hspace{0.5em}≃\hspace{0.5em}\AyaData{Int}

\setlength{\leftskip}{0cm}
\vspace{0.15cm}
}

How do we deal with elimination of this type?
\begin{observation}\label{obs:int-map}
Maps from a natural number to $X$ always provide the following information:
\begin{itemize}
  \item An instance $x : X$,
  \item A function $f : X \to X$.
\end{itemize}
This is because every map $h$ from a natural number to $X$ consists of:
\begin{itemize}
  \item $h(0)$, which is an instance of $X$,
  \item $h(n+1)$, which can use $h(n):X$ and constructs an instance of $X$,
  which is essentially a map from $X$ to $X$.
\end{itemize}
Maps from an integer, analogously, always provide the following information:
\begin{itemize}
  \item An instance $x : X$,
  \item An equivalence $e : X \simeq X$.
\end{itemize}
This is what we need to provide in the eliminator!
\end{observation}
But the above imaginary syntax is not supported by any known type theory
(which may also worth exploration!).
One way to make sense of this is to unfold the definition of equivalence,
which is discussed in detail in~\cite{Integers}.

\begin{observation}\label{obs:succ}
For the definition of $\mathbb{Z}_U$, we may apply \AyaData{PreInt} to \AyaConstructor{path}
to get the analogous thing of \AyaConstructor{succ} as in~\cref{obs:int-map}:

{
\vspace{0.15cm}
\RaggedRight
\setlength\parindent{0pt}
\setlength{\leftskip}{1cm}
\AyaKeyword{def}\hspace{0.5em}\AyaFn{succPath}\hspace{0.5em}\(:\)\hspace{0.5em}\AyaFn{Int}\hspace{0.5em}\AyaKeyword{↑}\hspace{0.5em}\AyaFn{=}\hspace{0.5em}\AyaFn{Int}\hspace{0.5em}\(\Rightarrow\)\hspace{0.5em}\AyaFn{ap}\hspace{0.5em}\AyaData{PreInt}\hspace{0.5em}\AyaConstructor{path}

\setlength{\leftskip}{0cm}
\vspace{0.15cm}
}

The $\uparrow$ symbol lifts the $=$ to a higher universe level,
so we can put types on the two sides of a path.\footnote{
Ideally, this should be inferred automatically, but we haven't implemented it yet.}
It is harmless in this particular case to just ignore it.
The design follows Conor McBride's proposal of implementing universe polymorphism,
studied in~\cite{McBrideUniv}.
\end{observation}

\begin{observation}
The equivalence in \cref{obs:succ} gives the successor and predecessor function on integers,
as well as the fact that they are inverse to each other.
\end{observation}

The other way is to think about Szumi Xie's trick.
When we want to define $n$ separate types, we can define one type \textit{indexed} by
an enumeration of $n$ constructors.
The enumeration is considered a ``mini-universe'',
as it \textit{governs} several types.

Now we want to define a type \AyaData{Int} with an equivalence on itself,
so instead we define a type indexed by a ``higher-enumeration'' \AyaData{U} with two constructors:
\begin{itemize}
  \item \AyaConstructor{Int} for the type we want to define,
  \item \AyaConstructor{path} of type $\AyaConstructor{\mathrm{Int}} = \AyaConstructor{\mathrm{Int}}$
    for the equivalence on the type itself.
    We work with path instead of the actual equivalence definition
    because only the former is supported by higher inductive types.
\end{itemize}
Then, the type indexed by \AyaData{U} (we call it \AyaData{PreInt})
will have some constructors that correspond
to the constructors of \AyaData{Int}. For \AyaConstructor{zero}, we add
\[
  \AyaConstructor{\mathrm{zero}_T} : \AyaData{\mathrm{PreInt}}~\AyaConstructor{T}
\]
And the good news is that we don't need to do anything for \AyaConstructor{succ},
because we can directly derive it from path induction on \AyaConstructor{path}.
Assuming the following operator for substitution and path reversal
(implementation omitted to avoid distraction):

{
\vspace{0.15cm}
\RaggedRight
\setlength\parindent{0pt}
\setlength{\leftskip}{1cm}
\AyaKeyword{def}\hspace{0.5em}\AyaFn{subst}\hspace{0.5em}\((\)\AyaLocalVar{P}\hspace{0.5em}\(:\)\hspace{0.5em}\AyaGeneralized{A}\hspace{0.5em}\(\to\)\hspace{0.5em}\AyaKeyword{Type}\()\)\hspace{0.5em}\(\{\)\AyaLocalVar{a}\hspace{0.5em}\AyaLocalVar{b}\hspace{0.5em}\(:\)\hspace{0.5em}\AyaGeneralized{A}\(\}\)\hspace{0.5em}\((\)\AyaLocalVar{p}\hspace{0.5em}\(:\)\hspace{0.5em}\AyaLocalVar{a}\hspace{0.5em}\AyaFn{=}\hspace{0.5em}\AyaLocalVar{b}\()\)\hspace{0.5em}\(:\)\hspace{0.5em}\AyaLocalVar{P}\((\)\AyaLocalVar{a}\()\)\hspace{0.5em}\(\to\)\hspace{0.5em}\AyaLocalVar{P}\((\)\AyaLocalVar{b}\()\)

\AyaKeyword{def}\hspace{0.5em}\AyaFn{sym}\hspace{0.5em}\(\{\)\AyaLocalVar{a}\hspace{0.5em}\AyaLocalVar{b}\hspace{0.5em}\(:\)\hspace{0.5em}\AyaGeneralized{A}\(\}\)\hspace{0.5em}\((\)\AyaLocalVar{p}\hspace{0.5em}\(:\)\hspace{0.5em}\AyaLocalVar{a}\hspace{0.5em}\AyaFn{=}\hspace{0.5em}\AyaLocalVar{b}\()\)\hspace{0.5em}\(:\)\hspace{0.5em}\AyaLocalVar{b}\hspace{0.5em}\AyaFn{=}\hspace{0.5em}\AyaLocalVar{a}

\setlength{\leftskip}{0cm}
\vspace{0.15cm}
}

We can define the functions \AyaFn{succ} and even \AyaFn{pred} as follows:

{
\vspace{0.15cm}
\RaggedRight
\setlength\parindent{0pt}
\setlength{\leftskip}{1cm}
\AyaKeyword{def}\hspace{0.5em}\AyaFn{succ}\hspace{0.5em}\(:\)\hspace{0.5em}\AyaFn{Int}\hspace{0.5em}\(\to\)\hspace{0.5em}\AyaFn{Int}\hspace{0.5em}\(\Rightarrow\)\hspace{0.5em}\AyaFn{subst}\hspace{0.5em}\AyaData{PreInt}\hspace{0.5em}\AyaConstructor{path}~\\
\AyaKeyword{def}\hspace{0.5em}\AyaFn{pred}\hspace{0.5em}\(:\)\hspace{0.5em}\AyaFn{Int}\hspace{0.5em}\(\to\)\hspace{0.5em}\AyaFn{Int}\hspace{0.5em}\(\Rightarrow\)\hspace{0.5em}\AyaFn{subst}\hspace{0.5em}\AyaData{PreInt}\hspace{0.5em}\((\)\AyaFn{sym}\hspace{0.5em}\AyaConstructor{path}\()\)

\setlength{\leftskip}{0cm}
\vspace{0.15cm}
}

Note that \AyaFn{succ} is just the underlying function of the map in~\cref{obs:succ}.
We may also show that they're inverse to each other,
because substitution commutes with path composition, say,
we have the following path equivalences:
\begin{align*}
  \AyaFn{\mathrm{subst}}~A~p~(\AyaFn{\mathrm{subst}}~A~q~u) &= \AyaFn{\mathrm{subst}}~A~(p\cdot q)~u\\
  \AyaFn{\mathrm{subst}}~A~\AyaFn{\mathrm{idp}}~u &= u
\end{align*}
These lemmas directly leads to the desired results.

\begin{remark}\label{rem:succt}
In cubical type theory one can try to write this:
\[
  \AyaConstructor{\mathrm{succ}_T} : (i:\mathbb{I}) \to \AyaData{\mathrm{PreInt}}~(\AyaConstructor{\mathrm{path}}~i)
\]
It is difficult to make sense of this thing at this point,
because the boundary information needs to be taken care of, and we do not yet know where
this should happen. This will be addressed later.
\end{remark}

\section{Relation to circle}
Observe that the definition \AyaData{T} is identical to the circle $S^1$,
so we can rewrite the above definition with some renamings:

{
\vspace{0.15cm}
\RaggedRight
\setlength\parindent{0pt}
\setlength{\leftskip}{1cm}
\AyaKeyword{open}\hspace{0.5em}\AyaKeyword{data}\hspace{0.5em}\AyaData{S¹}~\\
\(\mid\)\hspace{0.5em}\AyaConstructor{base}~\\
\(\mid\)\hspace{0.5em}\AyaConstructor{loop}\hspace{0.5em}\(:\)\hspace{0.5em}\AyaConstructor{base}\hspace{0.5em}\AyaFn{=}\hspace{0.5em}\AyaConstructor{base}~\\
~\\
\AyaKeyword{open}\hspace{0.5em}\AyaKeyword{data}\hspace{0.5em}\AyaData{Helix}\hspace{0.5em}\AyaData{S¹}~\\
\(\mid\)\hspace{0.5em}\AyaConstructor{base}\hspace{0.5em}\(\Rightarrow\)\hspace{0.5em}\AyaConstructor{zero}~\\
~\\
\AyaKeyword{def}\hspace{0.5em}\AyaFn{Int}\hspace{0.5em}\(\Rightarrow\)\hspace{0.5em}\AyaData{Helix}\hspace{0.5em}\AyaConstructor{base}

\setlength{\leftskip}{0cm}
\vspace{0.15cm}
}
\newcommand{\baseCon}{\AyaConstructor{\mathrm{base}}}

In essence, this definition of integers is directly related to the path space of $S^1$.
To make sense of this, we first refer to the famous theorem in homotopy type theory~\cite{Pi1S1}:
\[
  (\baseCon = \baseCon) \simeq \mathbb{Z}
\]
The proof is by showing that, for $x:S^1$,
the following type applied by $x$ is equivalent to the path space of $S^1$ based at $x$:

{
\vspace{0.15cm}
\RaggedRight
\setlength\parindent{0pt}
\setlength{\leftskip}{1cm}
\AyaKeyword{def}\hspace{0.5em}\AyaFn{Code}\hspace{0.5em}\AyaData{S¹}\hspace{0.5em}\(:\)\hspace{0.5em}\AyaKeyword{Type}~\\
\(\mid\)\hspace{0.5em}\AyaConstructor{base}\hspace{0.5em}\(\Rightarrow\)\hspace{0.5em}\AyaData{Z}~\\
\(\mid\)\hspace{0.5em}\AyaConstructor{loop}\hspace{0.5em}\AyaLocalVar{i}\hspace{0.5em}\(\Rightarrow\)\hspace{0.5em}\AyaFn{univalence}\hspace{0.5em}\AyaFn{succEquiv}\hspace{0.5em}\AyaLocalVar{i}

\setlength{\leftskip}{0cm}
\vspace{0.15cm}
}

And the two directions of the theorem are called \textit{encode} and \textit{decode} respectively:
\begin{align*}
  \mathrm{encode} &: (x:S^1) \to (\baseCon = x) \to \mathrm{Code}(x) \\
  \mathrm{decode} &: (x:S^1) \to \mathrm{Code}(x) \to (\baseCon = x)
\end{align*}

The fact that encode and decode are inverses of each other can be illustrated in the following way:
\begin{itemize}
  \item The base point corresponds to $\mathrm{Code}(\baseCon) \equiv \mathrm{Z}$:

  \begin{tikzpicture}[scale=.5]
    \draw (2,0) ellipse (1.5 and 0.5);
    \node[label=left:{\AyaConstructor{base}}] at ($(2,0)+(left:1.5 and 0.5)$) {\textbullet};
  
    \draw (2,1.5) ellipse (1.5 and 0.5);
    \node[label=left:{\AyaData{Z}}] at ($(2,1.5)+(left:1.5 and 0.5)$) {\textbullet};
  \end{tikzpicture}
  \item The loop corresponds to the path obtained from applying \AyaFn{succEquiv} to univalence:

  \begin{tikzpicture}[scale=.5]
    \draw (2,0) ellipse (1.5 and 0.5);
    \node[label=right:{\AyaConstructor{loop}~$i$}] at ($(2,0)+(right:1.5 and 0.5)$) {\textbullet};
    \node[label=left:{\AyaConstructor{base}}] at ($(2,0)+(left:1.5 and 0.5)$) {\textbullet};

    \draw (2,1.5) ellipse (1.5 and 0.5);
    \node[label=right:{\AyaFn{univalence}~\AyaFn{succEquiv}~$i$}] at ($(2,1.5)+(right:1.5 and 0.5)$) {\textbullet};
    \node[label=left:{\AyaData{Z}}] at ($(2,1.5)+(left:1.5 and 0.5)$) {\textbullet};
  \end{tikzpicture}
\end{itemize}

\begin{remark}
We have omitted the following definition for spatial considerations\footnote{
In fact, we haven't finish implementing univalence in Aya yet.}:
\begin{itemize}
  \item \AyaData{Z} is the type of integers. This is not to be confused with \AyaData{Int},
    which is a concrete definition given above.
    Here \AyaData{Z} can be any definition of integers, preferably a simple, intuitive one.
  \item \AyaFn{succ} is the successor function on \AyaData{Z}.
    Every integer type should have such a function.
  \item \AyaFn{succEquiv} is an equivalence between \AyaData{Z} and itself,
    given by \AyaFn{succ}.
  \item \AyaFn{univalence} takes an equivalence and returns a path.
\end{itemize}
Another purpose of this emission is to encourage readers to think about integers abstractly.

Also, Aya uses the cubical definition of path, which is similar to a function from the interval type.
That's also why the pattern for \AyaConstructor{loop} has a parameter.
Curious readers can refer to~\cite{CHM} for more details.
\end{remark}

\begin{observation}
In the definition of \AyaFn{Code}, we put \AyaData{Z} and an equivalence on itself
that corresponds to \AyaFn{succ} to the function bodies.
As observed in~\cref{obs:int-map}, this is the essential information we want on integers.

By encode-decode of $S^1$, we know that \AyaFn{Code} is equivalent to the path space of $S^1$.
So, if somehow we can make the path space of $S^1$ as the \textit{definition} of integers,
then it will be really nice.
\end{observation}

\section{Fording indexed families}

Henry Ford is quoted for having said that
``Any color the customer wants, as long as it's black.''

An analogous method is used to encode indexed families with a built-in identity type:
suppose we have a family $F : A \to \mathcal U$ with a constructor
$c : X \to F~a$ for some expression $a : A$ (usually in neutral form),
Fording turns it into the following:
\[
  c_{\mathrm{Ford}} : (a' : A) \to (a = a') \to X \to F~a'
\]
And now $F$ is no longer \textit{indexed}.

\begin{example}
Consider the following indexed family:

{
\vspace{0.15cm}
\RaggedRight
\setlength\parindent{0pt}
\setlength{\leftskip}{1cm}
\AyaKeyword{open}\hspace{0.5em}\AyaKeyword{data}\hspace{0.5em}\AyaData{Bool}~\\
\(\mid\)\hspace{0.5em}\AyaConstructor{true}~\\
\(\mid\)\hspace{0.5em}\AyaConstructor{false}~\\
~\\
\AyaKeyword{open}\hspace{0.5em}\AyaKeyword{data}\hspace{0.5em}\AyaData{So}\hspace{0.5em}\AyaData{Bool}~\\
\(\mid\)\hspace{0.5em}\AyaConstructor{true}\hspace{0.5em}\(\Rightarrow\)\hspace{0.5em}\AyaConstructor{oh}

\setlength{\leftskip}{0cm}
\vspace{0.15cm}
}
We have:
\begin{itemize}
  \item $\AyaConstructor{\mathrm{oh}}:\AyaData{\mathrm{So}}~\AyaConstructor{\mathrm{true}}$
    is the unique inhabitant,
  \item $\AyaData{\mathrm{So}}~\AyaConstructor{\mathrm{false}}$
    is the empty type.
\end{itemize}
This type is extensively used in the Idris language (as \textsf{Data.So}\footnote{
\url{https://www.idris-lang.org/docs/idris2/current/base_docs/docs/Data.So.html}}),
where it lifts any Boolean-valued test to a type-level proposition.
Fording it gives us the following type:

{
\vspace{0.15cm}
\RaggedRight
\setlength\parindent{0pt}
\setlength{\leftskip}{1cm}
\AyaKeyword{open}\hspace{0.5em}\AyaKeyword{data}\hspace{0.5em}\AyaData{So}\hspace{0.5em}\((\)\AyaLocalVar{b}\hspace{0.5em}\(:\)\hspace{0.5em}\AyaData{Bool}\()\)~\\
\(\mid\)\hspace{0.5em}\AyaConstructor{oh}\hspace{0.5em}\((\)\AyaLocalVar{b}\hspace{0.5em}\AyaFn{=}\hspace{0.5em}\AyaConstructor{true}\()\)

\setlength{\leftskip}{0cm}
\vspace{0.15cm}
}
Then we can see directly that $\AyaData{\mathrm{So}}~b$ is
nothing but a wrapper of $b=\AyaConstructor{\mathrm{true}}$.
\end{example}

\begin{exercise}
Ford the standard definition of finite sets (called \textsf{Fin} in Idris) and the sized vector type.
\end{exercise}

Fording helps us get rid of index unification in pattern matching.
Suppose we want to split on the type $F~b$ for some $b : A$,
then if we write the pattern $c~x$, it must be the case that $b$ and $a$ can be
\textit{unified}, otherwise this pattern matching will be \textit{rejected}.
In case $b$ is a function call that stuck on some argument,
this unification is very unlikely to succeed.

\begin{example}
We cannot perform pattern matching on a type $\AyaData{\mathrm{So}}~(\mathrm{isEmpty}~x)$
where $x : \mathrm{List}~A$ is a variable.
\end{example}

\begin{remark}
In traditional dependent pattern matching~\cite{DepPM},
writing \AyaConstructor{oh} as a pattern for the type $\AyaData{\mathrm{So}}~x$
(when $x$ is a variable) will \textit{rewrite} $x$ as $\AyaConstructor{\mathrm{true}}$
in the right-hand side of the clause:

{
\vspace{0.15cm}
\RaggedRight
\setlength\parindent{0pt}
\setlength{\leftskip}{1cm}
\AyaFn{f}\hspace{0.5em}:\hspace{0.5em}\((\)\AyaLocalVar{x}\hspace{0.5em}:\hspace{0.5em}\AyaData{Bool}\()\)\hspace{0.5em}\(\to\)\hspace{0.5em}\AyaData{So}\hspace{0.5em}\AyaLocalVar{x}\hspace{0.5em}\(\to\)\hspace{0.5em}\AyaLocalVar{x}\hspace{0.5em}=\hspace{0.5em}\AyaConstructor{true}~\\
\AyaFn{f}\hspace{0.5em}\(\_\)\hspace{0.5em}\AyaConstructor{oh}\hspace{0.5em}=\hspace{0.5em}\AyaConstructor{refl}

\setlength{\leftskip}{0cm}
\vspace{0.15cm}
}
We assume $\AyaConstructor{\mathrm{refl}}$ to be the constructor of the inductively-defined identity type.

This is a desirable feature because it matches the intuition that ``only when $x$ is true can
we construct \AyaConstructor{oh} as a proof of $\AyaData{\mathrm{So}}~x$''.

After Fording, we can do the same by writing $\AyaConstructor{\mathrm{oh}}~\AyaConstructor{\mathrm{refl}}$:

{
\vspace{0.15cm}
\RaggedRight
\setlength\parindent{0pt}
\setlength{\leftskip}{1cm}
\AyaFn{f}\hspace{0.5em}:\hspace{0.5em}\((\)\AyaLocalVar{x}\hspace{0.5em}:\hspace{0.5em}\AyaData{Bool}\()\)\hspace{0.5em}\(\to\)\hspace{0.5em}\AyaData{So}\hspace{0.5em}\AyaLocalVar{x}\hspace{0.5em}\(\to\)\hspace{0.5em}\AyaLocalVar{x}\hspace{0.5em}=\hspace{0.5em}\AyaConstructor{true}~\\
\AyaFn{f}\hspace{0.5em}\(\_\)\hspace{0.5em}\((\)\AyaConstructor{oh}\hspace{0.5em}\AyaConstructor{refl}\()\)\hspace{0.5em}=\hspace{0.5em}\AyaConstructor{refl}

\setlength{\leftskip}{0cm}
\vspace{0.15cm}
}
But what's good about Fording is that we can \textit{delay} this rewriting to the moment
when we actually need it, especially in case $x$ is not a variable.
In fact, in the above case, we don't even do any dependent pattern matching:

{
\vspace{0.15cm}
\RaggedRight
\setlength\parindent{0pt}
\setlength{\leftskip}{1cm}
\AyaFn{f}\hspace{0.5em}:\hspace{0.5em}\((\)\AyaLocalVar{x}\hspace{0.5em}:\hspace{0.5em}\AyaData{Bool}\()\)\hspace{0.5em}\(\to\)\hspace{0.5em}\AyaData{So}\hspace{0.5em}\AyaLocalVar{x}\hspace{0.5em}\(\to\)\hspace{0.5em}\AyaLocalVar{x}\hspace{0.5em}=\hspace{0.5em}\AyaConstructor{true}~\\
\AyaFn{f}\hspace{0.5em}\(\_\)\hspace{0.5em}\((\)\AyaConstructor{oh}\hspace{0.5em}\AyaLocalVar{p}\()\)\hspace{0.5em}=\hspace{0.5em}\AyaLocalVar{p}

\setlength{\leftskip}{0cm}
\vspace{0.15cm}
}
This also prevents over-eager rewriting of variables, which can potentially lead to unreadable
compiler output when doing type-directed development, or slow down proof search.
\end{remark}

Now, let's apply this to the definition of integers:

{
\vspace{0.15cm}
\RaggedRight
\setlength\parindent{0pt}
\setlength{\leftskip}{1cm}
\AyaKeyword{open}\hspace{0.5em}\AyaKeyword{data}\hspace{0.5em}\AyaData{S¹}~\\
\(\mid\)\hspace{0.5em}\AyaConstructor{base}~\\
\(\mid\)\hspace{0.5em}\AyaConstructor{loop}\hspace{0.5em}\(:\)\hspace{0.5em}\AyaConstructor{base}\hspace{0.5em}\AyaFn{=}\hspace{0.5em}\AyaConstructor{base}~\\
~\\
\AyaKeyword{open}\hspace{0.5em}\AyaKeyword{data}\hspace{0.5em}\AyaData{Helix}\hspace{0.5em}\((\)\AyaLocalVar{x}\hspace{0.5em}\(:\)\hspace{0.5em}\AyaData{S¹}\()\)~\\
\(\mid\)\hspace{0.5em}\AyaConstructor{zero}\hspace{0.5em}\((\)\AyaLocalVar{x}\hspace{0.5em}\AyaFn{=}\hspace{0.5em}\AyaConstructor{base}\()\)~\\
~\\
\AyaKeyword{def}\hspace{0.5em}\AyaFn{Int}\hspace{0.5em}\(\Rightarrow\)\hspace{0.5em}\AyaData{Helix}\hspace{0.5em}\AyaConstructor{base}

\setlength{\leftskip}{0cm}
\vspace{0.15cm}
}

\begin{observation}
The type \AyaData{Helix} is simply a wrapper
of the path space of $S^1$ based at a point given by the parameter $x:S^1$.
\end{observation}

\begin{observation}
The type \AyaData{Int}, when we look through the wrapper \AyaData{Helix},
is an alias to the type $\baseCon = \baseCon$.
\end{observation}

The same idea applies to any inhabited type. Consider such a case $\braket{H:\mathcal U, x:H}$.
The path space of $H$ based at $x:H$ can be defined as $P : H \to \mathcal U$
with a constructor $c : P~x$.

Fording also reveals the mystery behind the constructor
$\AyaConstructor{\mathrm{succ}_T}$ in~\cref{rem:succt} by expanding it to:

{
\vspace{0.15cm}
\RaggedRight
\setlength\parindent{0pt}
\setlength{\leftskip}{1cm}
\AyaKeyword{open}\hspace{0.5em}\AyaKeyword{data}\hspace{0.5em}\AyaData{Helix'}\hspace{0.5em}\((\)\AyaLocalVar{x}\hspace{0.5em}\(:\)\hspace{0.5em}\AyaData{S¹}\()\)~\\
\(\mid\)\hspace{0.5em}\AyaConstructor{zero}\hspace{0.5em}\((\)\AyaLocalVar{x}\hspace{0.5em}\AyaFn{=}\hspace{0.5em}\AyaConstructor{base}\()\)~\\
\(\mid\)\hspace{0.5em}\AyaConstructor{succ}\hspace{0.5em}\((\)\AyaLocalVar{i}\hspace{0.5em}\(:\)\hspace{0.5em}\AyaPrimitive{I}\()\)\hspace{0.5em}\((\)\AyaLocalVar{x}\hspace{0.5em}\AyaFn{=}\hspace{0.5em}\AyaConstructor{loop}\hspace{0.5em}\AyaLocalVar{i}\()\)

\setlength{\leftskip}{0cm}
\vspace{0.15cm}
}

And we know how to read this type in cubical type theory: it's a square!
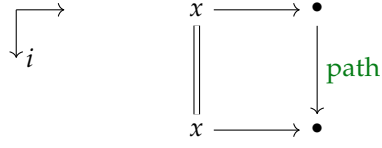
\carloCTikZ{
\carloCoordSq{}{$i$}
\node(0) at (0, 0) {$x$};
\node(1) at (1, 0) {\textbullet};
\node(2) at (0, 1) {$x$};
\node(3) at (1, 1) {\textbullet};
\draw[->] (0) -- (1);
\draw[->] (2) -- (3);
\draw[equals arrow] (0) -- (2);
\draw[->] (1) -- (3) node [midway,right] {$\AyaConstructor{\mathrm{path}}$};
}
So the constructor is just parameterized by a square in a peculiar way.

\section{Further discussion on syntax}

In~\cref{obs:int-map}, we have given an example of type that is \textit{freely generated}
by a constructor and an equivalence on the type, where the equivalence is a constructor
equipped with the property of being an equivalence
(called \AyaFn{isEquiv} in various formalization projects).

Even though this can only be expressed by expanding the definition of equivalence and will
end up giving us loads of constructors, it is a powerful notation for informal writings.

\begin{example}
A real-world use case is in~\cite[\S 2]{PathPO},
which cites a result from~\cite{PathHIT} that certain path space of pushout is
freely generated by reflexivity and an equivalence.
The pushout discussed in the paper is defined for a span,
which includes $A : \mathcal U, B : \mathcal U, R : A \to B \to \mathcal U$,
and we write $A +_R B$ for the pushout, and
\newcommand{\inlCon}{\AyaConstructor{\mathrm{inl}}}
\newcommand{\inrCon}{\AyaConstructor{\mathrm{inr}}}
\begin{align*}
  \inlCon &: A \to A +_R B \\
  \inrCon &: B \to A +_R B
\end{align*}
as two of its constructors.
The path spaces of interest are:
\[
  \inlCon(a_0) = \inlCon(a) \quad \inlCon(a_0) = \inrCon(b) \quad \inrCon(b) = \inrCon(b')
\]
And the constructors for these types are:

{
\vspace{0.15cm}
\RaggedRight
\setlength\parindent{0pt}
\setlength{\leftskip}{1cm}
\AyaKeyword{data}\hspace{0.5em}\AyaData{PathSpaces}\hspace{0.5em}\((\)\AyaLocalVar{A}\hspace{0.5em}\AyaLocalVar{B}\hspace{0.5em}:\hspace{0.5em}\AyaKeyword{Type}\()\)\hspace{0.5em}\((\)\AyaLocalVar{R}\hspace{0.5em}:\hspace{0.5em}\AyaLocalVar{A}\hspace{0.5em}\(\to\)\hspace{0.5em}\AyaLocalVar{B}\hspace{0.5em}\(\to\)\hspace{0.5em}\AyaKeyword{Type}\()\)\hspace{0.5em}\AyaKeyword{where}~\\
\hspace{1.0em}\AyaConstructor{refl}\hspace{0.5em}:\hspace{0.5em}\((\)\AyaLocalVar{a₀}\hspace{0.5em}:\hspace{0.5em}\AyaLocalVar{A}\()\)\hspace{0.5em}\(\to\)\hspace{0.5em}\AyaConstructor{inl}\((\)\AyaLocalVar{a₀}\()\)\hspace{0.5em}=\hspace{0.5em}\AyaConstructor{inl}\((\)\AyaLocalVar{a₀}\()\)~\\
\hspace{1.0em}\AyaConstructor{eq}\hspace{0.5em}:\hspace{0.5em}\((\)\AyaLocalVar{a₀}\hspace{0.5em}\AyaLocalVar{a}\hspace{0.5em}:\hspace{0.5em}\AyaLocalVar{A}\()\)\hspace{0.5em}\(\to\)\hspace{0.5em}\((\)\AyaLocalVar{b}\hspace{0.5em}:\hspace{0.5em}\AyaLocalVar{B}\()\)\hspace{0.5em}\(\to\)\hspace{0.5em}\AyaLocalVar{R}\hspace{0.5em}\AyaLocalVar{a}\hspace{0.5em}\AyaLocalVar{b}\hspace{0.5em}\(\to\)~\\
\hspace{1.0em}\hspace{1.0em}\hspace{1.0em}\((\)\AyaConstructor{inl}\((\)\AyaLocalVar{a₀}\()\)\hspace{0.5em}=\hspace{0.5em}\AyaConstructor{inl}\((\)\AyaLocalVar{a}\()\)\()\)\hspace{0.5em}≃\hspace{0.5em}\((\)\AyaConstructor{inl}\((\)\AyaLocalVar{a₀}\()\)\hspace{0.5em}=\hspace{0.5em}\AyaConstructor{inr}\((\)\AyaLocalVar{b}\()\)\()\)

\setlength{\leftskip}{0cm}
\vspace{0.15cm}
}
Without homotopy type theory,
it should be quite difficult to describe this path space,
and even if it's written by someone, it's less likely to be understood by others
compared to this one.
\end{example}

\printbibliography
\end{document}